\documentclass[12pt]{article}
\usepackage{graphicx}
\usepackage{amsmath}
\usepackage{txfonts}
\begin{document}
\newcommand{\rmn}[1] {{\rm #1}}
\newcommand{\Zsolar}{\mbox{${\; {\rm {\sun}}}$}}
\newcommand{\ha}{\hbox{H$\alpha$}}
\newcommand{\oii}{\hbox{[O\,{\sc ii}]}}
\newcommand{\neiii}{\hbox{[Ne\,{\sc iii}]}}
\newcommand{\siii}{\hbox{[S\,{\sc iii}]}}
\newcommand{\ariii}{\hbox{[Ar\,{\sc iii}]}}
\newcommand{\nii}{\hbox{[N\,{\sc ii}]}}
\newcommand{\sii}{\hbox{[S\,{\sc ii}]}}
\newcommand{\oiii}{\hbox{[O\,{\sc iii}]}}
\newcommand{\hb}{\hbox{H$\beta$}}
\newcommand{\hg}{\hbox{H$\gamma$}}
\newcommand{\hd}{\hbox{H$\delta$}}
\newcommand{\hi}{\hbox{H\,{\sc i}}}
\newcommand{\hii}{\hbox{H\,{\sc ii}}}
\newcommand{\heii}{\hbox{He\,{\sc ii}}}
\newcommand{\etal}{\hbox{et\thinspace al.\ }}
\newcommand{\oiiha}{\hbox{[O\,{\sc ii}]/H$\alpha$}}
\newcommand{\fha}{\hbox{$F_{{\rm H}\alpha}$}}
\newcommand{\fhb}{\hbox{$F_{{\rm H}\beta}$}}
\newcommand{\foii}{\hbox{$F_{\rm [O\,{\sc II}]}$}}
\newcommand{\micron}{\hbox{$\mu$m}}
\newcommand{\zoh}{\hbox{$12\,+\,{\rm log(O/H)}$}}
\newcommand{\teoii} {\hbox{$T_e{\rm (O\,{\sc II})}$}}
\newcommand{\tenii} {\hbox{$T_e{\rm (N\,{\sc II})}$}}
\newcommand{\teoiii}{\hbox{$T_e{\rm (O\,{\sc III})}$}}
\newcommand{\mnras}{MNRAS}
\newcommand{\apj}{ApJ}
\newcommand{\apjl}{ApJL}
\newcommand{\pasp}{PASP}
\newcommand{\aap}{A\&A}
\newcommand{\araa}{ARA\&A}
\newcommand{\apjs}{ApJS}
\newcommand{\aj}{AJ}
\newcommand{\jrasc}{JRASC}
 {\center {\large STAR FORMATION RATE INDICATORS IN WIDE-FIELD INFRARED SURVEY Preliminary Release\\}}

 Fei Shi, {\it Department of Basic Science, North China
institute of spaceflight engineering , Hebei 065000, PR
China,e-mail: fshi@bao.ac.cn\\}

Xu Kong, {\it  Center of Astrophysics, University of Science and
Technology of China, Jinzhai Road 96, Hefei 230026, China.\\}

 James Wicker, {\it  National Astronomical Observatories,
Chinese Academy of Sciences, 20A Datun Road, Chaoyang District,
Beijing 100012, PR
              China\\}

Yang Chen, {\it  Center of Astrophysics, University of Science and
Technology of China, Jinzhai Road 96, Hefei 230026, China.\\}

Zi-Qiang, Gong, {\it Department of Basic Science, North China
institute of spaceflight engineering , Hebei 065000, PR
China\\}

Dong-Xin, Fan, {\it Department of physics and Electronics, Guangxi Teachers
Education University, Nanning 530001, China.\\}

 {\bf Abstract.} With the goal of investigating the degree to which the MIR
luminosity in the WIDE-FIELD INFRARED SURVEY (WISE) traces the SFR,
we analyze  3.4, 4.6, 12 and  22 $\mu$m data  in a sample of $\sim$
140,000 star-forming galaxies or star-forming regions {\bf covering a wide range in metallicity
$7.66<\zoh<9.46$, in redshift less than 0.4}. These star-forming galaxies or star-forming regions are selected
by matching the WISE Preliminary Release Catalog with the star-forming
galaxy Catalog in SDSS DR8 provided by
JHU/MPA \footnote{http://www.sdss3.org/dr8/spectro/galspec.php.}.
We study the
relationship between the luminosity at 3.4, 4.6, 12 and 22 $\mu$m from
WISE and H$\alpha$ luminosity in SDSS DR8. From these comparisons, we
derive reference SFR indicators for use in our analysis. Linear correlations between SFR and the 3.4, 4.6, 12 and 22 $\mu$m
luminosity are found, and calibrations of SFRs based on L(3.4), L(4.6),
L(12) and L(22$\mu$m) are proposed. The calibrations hold for galaxies
with verified spectral observations. The dispersion in the relation
between 3.4, 4.6, 12 and 22 $\mu$m luminosity and SFR relates to the
galaxy's properties, such as 4000$\AA$ break and galaxy color.\\


 \pagebreak

\def\baselinestretch{1.66}
\large
\normalsize

%{\center{\bf 1. Introduction\\}}
\section{Introduction}
 The star formation rate (SFR) is a crucial parameter to characterize
the star formation history of galaxies.
To calculate the SFR reliably, extensive efforts have been made to
derive SFR indicators at various wavelengths, including radio, infrared
 (IR), ultraviolet (UV), optical spectral lines and continuum
(Kennicutt 1998).
Among these wavelengths, SFR indicators at the infrared (IR) band have
attracted more attention in recent years because of the high sensitivity
and high angular resolution data provided by the Spitzer Space Telescope.
As a result, the general correlation between infrared luminosity and SFR
has been found and calibrated ( Calzetti et al. 2005, 2007, 2009, 2010,
AlonsoHerrero et al. 2006,  Schmitt et al. 2006,
Kennicutt et al. 2007, Persic et al. 2007, Rosa et al. 2007, Salim et al.
2007, Bigiel et al. 2008, Rieke et al. 2009).

During last twenty years, the calibrations of these correlations are
mainly focused on the relation between the total luminosity in the IR
band ($L_{\rm TIR}$) and SFR because of dust heating in  the wide IR
band.
As a result, the monochromatic SFR indicators based on a single band
measurement from galaxies are ignored to some extent.
Calculation of $L_{\rm TIR}$ requires models for the infrared spectral
energy distribution (SED) of star-forming galaxies (Chary \& Elbaz 2001,
 Dale \& Helou 2002, Lagache et al.2003, Marcillac et al.2006, Noll et
al.2009), but these models usually suffer from small galaxy sample size
and limited sensitivities from surveys such as Infrared Astronomical
Satellite (IRAS) and Infrared Space Observatory (ISO).

 Because of the shortcomings of the $L_{\rm TIR}$, studies of
monochromatic SFR indicators based on the single band measurement from
 galaxies have experienced a new resurgence, such as the emission
detected in the 8~$\mu$m and 24~$\mu$m Spitzer bands. This emission has
been analyzed by a number of authors (Calzetti et al. 2005, 2007, 2010,
AlonsoHerrero et al. 2006, Perez--Gonzalez et al.2006, Rellano et al.
2007, Salim et al. 2007, Rieke et al. 2009), but the sample size of
Spitzer is  small and the sensitivity  is  also limited.

In a word, the calibrations based on the relation between the total
luminosity and SFR are still problematic because of  the limitation of
sensitivity and size of sample. This dearth of infrared data for normal
star-forming galaxies is largely a consequence of prior instrumental
limitations

The Wide-field Infrared Survey Explorer (WISE, Wright et al. 2010) will
 map the entire sky with 5 $\sigma$ point source sensitivities better
than 0.08, 0.11, 1 and 6 mJy at wavelengths of 3.4, 4.6, 12 \& 22
$\mu$m , which is 3 to 6 orders of magnitude more sensitive than
previous surveys.
For example, WISE is achieving 100 times better sensitivity than IRAS
in the 12 $\mu$m band. As an all-sky survey, WISE will finally return
data about over 500 million objects, so it provides us a large sample
of star-forming galaxies.
The improved sensitivity and large sample size make it suitable for
studying the evolution of galaxies.

The WISE Preliminary Release has been available to the astronomical
community since April 14, 2011 and contains the attributes for
257,310,278 objects observed during the first 105 days of the survey.
The data presented here were processed with initial calibrations and
reduction algorithms of the WISE pipeline derived from early survey
data (Cutri et al. 2011).

This paper is organized as follows. Based on the WISE Preliminary Data
Release, we present a sample to derive our SFR index calibration
(Sect. 2). In Sect. 3, we study the correlation between SFR and the
luminosity at 3.4, 4.6, 12 \& 22 $\mu$m and calibrate the 3.4, 4.6, 12
\& 22 $\mu$m SFR index. In Sect.4, we study the origin of dispersion
between SFR and the luminosity at 3.4, 4.6, 12 \& 22 $\mu$m.
 In Sect. 5, we summarize the calibration result and conclude this
paper. Throughout the paper, we adopt a value of the $\Omega_M$ = 0.27
and $\Omega_\Lambda$ = 0.73.

%{\center{\bf 2. Data sample\\}}\label{sam}
\section{Data sample}\label{sam}
The preferred method for determining SFR in star-forming
galaxies is obtained from the luminosity at some wavelengths.
WISE is surveying the entire sky at wavelengths of 3.4, 4.6, 12 \& 22 $\mu$m{\bf((W1 through W4, respectively)},
so we will study the correlation the SFR and the luminosity at 3.4, 4.6, 12 \& 22 $\mu$m and calibrate them.
In this paper, {\bf the  adopted SFR values are provided by
JHU/MPA, determined from the method in Brinchmann et al.(2004) which
combines emission line measurements from within the fiber where possible and aperture corrections are done
by fitting models of Gallazzi et al (2005), Salim et al (2007) to the photometry outside the fibre.
The luminosities L at 3.4, 4.6, 12 \& 22 $\mu$m are calculated  based on
red-shift and the "raw" source flux measured in the profile-fit photometry: 
 \begin{displaymath}
     Flux = \frac{L}{4 \pi \times  D^2} \nonumber
  \end{displaymath}
, where D is the luminosity distance derived from the red-shift. }

 We match the WISE Preliminary Release Catalog  with the star-forming galaxy catalog in SDSS
   DR 8 provided by MPA. The star-forming galaxy is selected by requiring the rigorous great circle arc distances
   between the galaxy of SDSS and WISE Preliminary Release Catalog
   to be less than the 6.1" which is the angular resolution of wavelength 1.
After converting digital numbers to Jy using the PSF-fit photometry (Cutri et al. 2011),
 we made the internal reddening correction for the flux of all the emission
lines, using the two strongest Balmer lines, \ha/\hb\ and
the effective absorption curve
$\tau_\lambda=\tau_V(\lambda/5500{\rm\AA})^{-0.7}$, which was
introduced by Charlot \& Fall (2000). Then, we made use of the spectral
diagnostic diagrams from Kauffmann et al. (2003) to classify galaxies
as star-forming galaxies, active galactic nuclei (AGN),
or unclassified.  To reduce systematic and random errors,
our galaxy samples are limited by the requirement that
the "raw"  source flux measured photometry is always larger than
three times of the uncertainty in the "raw" source flux measurement in profile-fit photometry.

In total, $\sim$140,000 star-forming galaxies are adopted in our sample.

\section{Star Formation Rates Calibrator}\label{sfr}

Hydrogen recombination line fluxes have been used very extensively
to estimate the SFR, since they are proportional to the
number of photons produced by the hot stars, which is in turn
proportional to their birthrate. Most applications of this method
have been based on measurements of luminosity from the H$\alpha$ line 
(Kong 2004).

Star-forming regions tend to be dusty and the dust absorption
cross-section peaks in the UV, and then is
reprocessed by dust and emerges beyond a few $\mu$m. As a result,
the luminosity at a few $\mu$m are reliable SFR indicators.
This process is restricted by the complex physical conditions, such as
not all of the luminous energy produced by recently formed stars is
re-processed by dust in the infrared depending on dust amount and
evolved non-star-forming stellar populations also heat the dust that then
emits in the FIR, etc{\bf(Calzetti et al. 2010, and references therein)}. So it is necessary to check whether the IR luminosity
can be reliable SFR indicators.

To check whether the luminosities at 3.4, 4.6, 12 \& 22 $\mu$m are reliable
SFR indicators, we show the relation between H$\alpha$ luminosity and the luminosity
at 3.4, 4.6, 12 \& 22 $\mu$m for our data sample in Fig. 1.
Apart from a few outliers in the  sample, all the objects merge into a
relatively tight, linear, and steep sequence, which gives the strong evidence that
the luminosities at 3.4, 4.6, 12 \& 22 $\mu$m are good SFR indicators just like the H$\alpha$ line.

%\begin{figure*}%f2
%\includegraphics[width=11.6cm]{test.eps}
%\caption {Shown in greyscale...}
%\label{cl1018}
%\end{figure*}

\begin{figure*}
   \centering
   \includegraphics[scale=0.7,angle=-90]{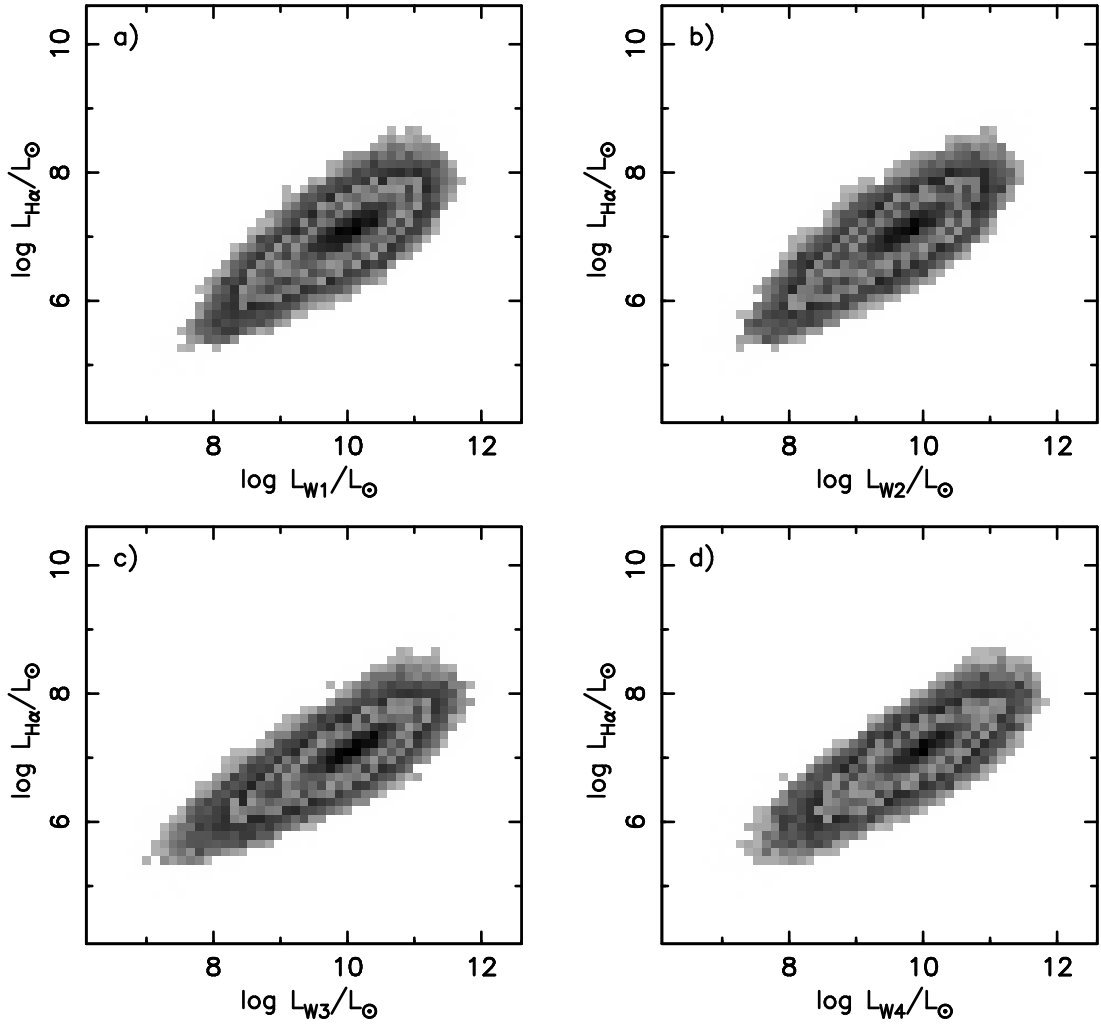}
   \caption{The relation between H$\alpha$ luminosity and the luminosity at W1 through W4, respectively, for our data sample.
   The luminosity is in units of solar luminosity.}
              \label{Fig1}%
    \end{figure*}

From the relation between SFR and the luminosity at 3.4, 4.6, 12 \& 22 $\mu$m
for our data sample in Fig. 2, we can define a new SFR
calibration. The observed distribution of all the points in
this figure is linear least-square-fitted by the expression
given as the dashed line in Fig. 2.

%\begin{eqnarray}
%\lg(SFR) = -7.90  + 0.80  \times \lg(L3.4\mu m)  \
%\end{eqnarray}

%\begin{eqnarray}
%\lg(SFR) = -6.60 + 0.67  \times \lg(L4.6\mu m)  \
%\end{eqnarray}

%\begin{eqnarray}
%\lg(SFR) = -7.50 + 0.78 \times \lg(L12\mu m)  \
%\end{eqnarray}

%\begin{eqnarray}
%\lg(SFR) = -6.90  + 0.70  \times \lg(L22\mu m)  \
%\end{eqnarray}

It should be noted that our SFR calibration holds only for the galaxies with authentic spectral observation.
Only galaxies which have reasonable values of oxygen abundance
come into Fig. 2($\sim$70,000 galaxies). All these galaxies have high quality spectral observations.
 If we plot the whole sample into Fig. 2, the dependencies of  SFR on
 the luminosity at 3.4, 4.6, 12 \& 22 $\mu$m will worsen.

In the next section, these fit's residuals will be discussed.

\begin{figure*}
   \centering
   \includegraphics[scale=0.7,angle=-90]{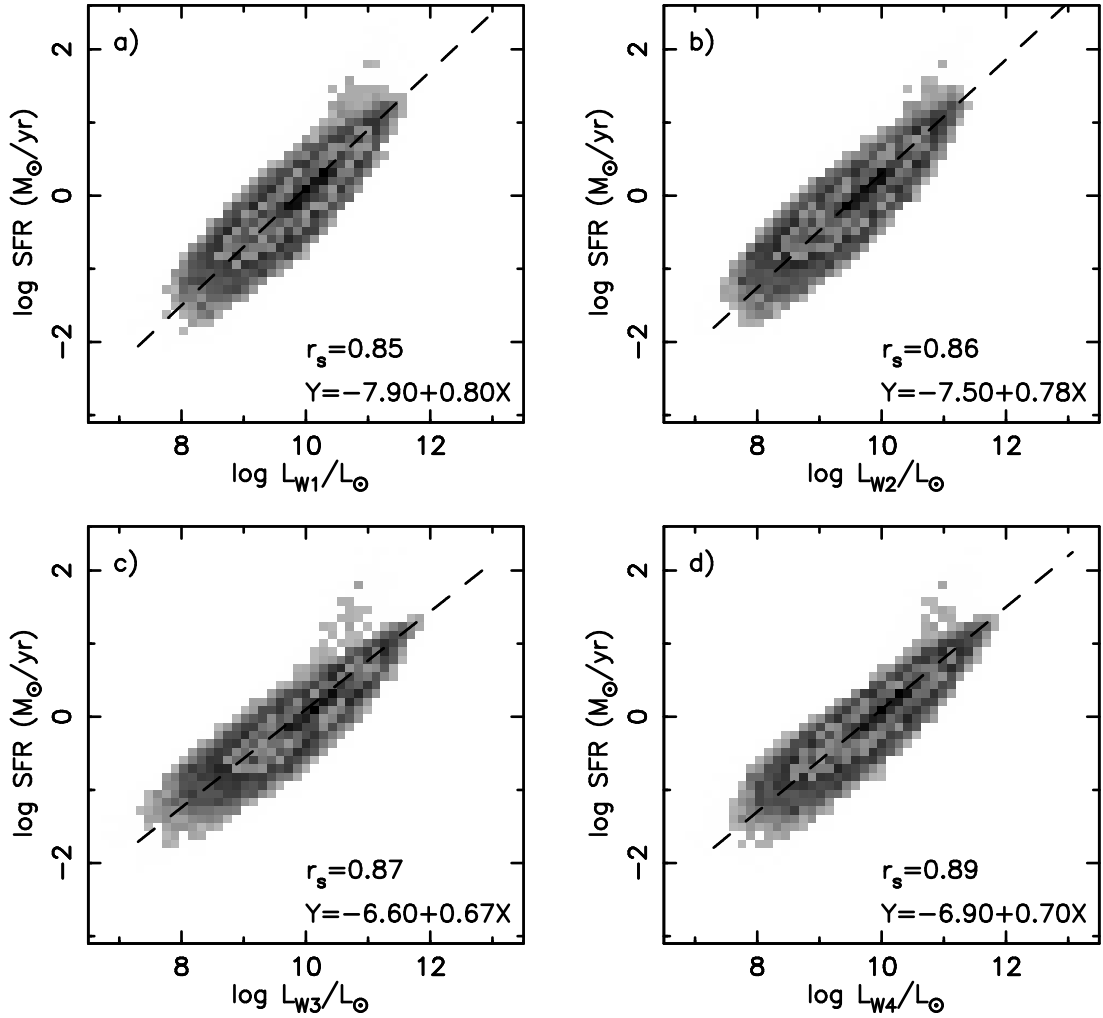}
   \caption{The relation between SFR and the luminosity at 3.4, 4.6, 12 \& 22 $\mu$m for our data sample. The black lines denote
the best-fit function. X and Y are luminosity and SFR, respectively. The SFR is in units of M$_{sun}$/yr. The luminosity is in units of solar luminosity. 
$r_s$ is the standard deviation of the fit.}
  \label{Fig2}%
    \end{figure*}

\section{The origin of the dispersion}

This close correlation between SFR and the luminosity can be understood by
the knowledge of the domination of the young star at the wavelengths of 12 \& 22 $\mu$m.
{\bf The contribution from the evolved non-star-forming stellar populations
at the wavelengths of 12 \& 22 $\mu$m are negligible for most star-forming galaxies in our sample .}
But for the wavelengths of  3.4, 4.6 $\mu$m, it is  established that the contribution
 by  the old star populations can be non-negligible in these wavelengths. The strong correlation
for the wavelength of  3.4, 4.6 $\mu$m in our data sample gives strong evidence
 that the contribution of young stellar populations are still dominate at the wavelength of  3.4, 4.6 $\mu$m.

%%%%% Check this point

\begin{figure*}
   \centering
   \includegraphics[scale=0.6,angle=-90]{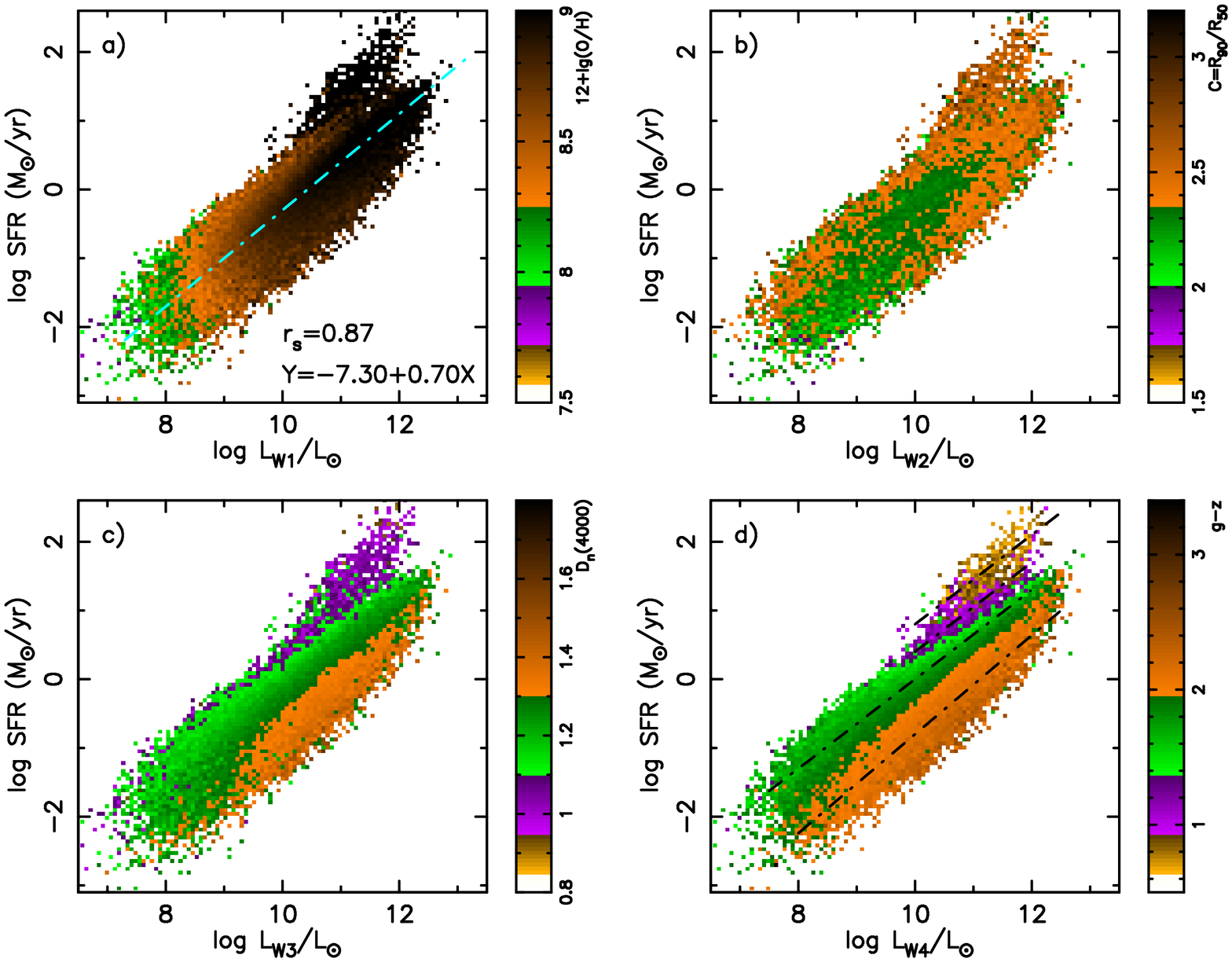}
   \caption{The relation between SFR and the luminosity  for our data sample.
   The sample in a) is subdivided into different oxygen abundances. 
   X and Y are luminosity and SFR, respectively. $r_s$ is the standard deviation of the fit.
   The sample in b) is subdivided into different concentration values.
   The sample in c) is subdivided into different 4000$\AA$ breaks.
   The sample in d) is subdivided into different g-z colors.
   The luminosity is in units of solar luminosity.}
              \label{Fig3}%
    \end{figure*}

%\subsection{The origin of the dispersion}

Although the SFR is closely related to the luminosity at 3.4, 4.6, 12 \& 22 $\mu$m,
the dispersion of the relation is significant. Furthermore,
the scatter does not show significant change for increasing wavelength,
but appears to increase for increasing luminosity (Fig. 2).

\subsection{Metallicity and Concentration index}

To further investigate these trends, we subdivide
the sample into sub-samples with different oxygen abundances (Fig. 3a).
The luminosity-SFR relation at low oxygen abundance is displaced to lower SFR and lower luminosity.
We can explain this because the majority of low oxygen abundance galaxies have less star forming
events, therefore they have relatively lower SFR.

The behavior that the higher oxygen abundance galaxies tends to have higher luminosity (mass) has been
studied by many authors (Shi et al. 2005, and references therein ). With the luminosity and metallicity
correlation, the most straightforward interpretation
 is that more massive galaxies form fractionally more stars in a Hubble time (higher luminosity)
 than low-mass galaxies, and then have higher metallicity.

The galaxies with the concentration
value $C > 2.6$ are mostly early type galaxies whereas
late type galaxies have $C < 2.6$ (Kauffmann et al.2003). It is well known that early
type galaxies are dominated by old/small mass stars and late
type galaxies are dominated by young/massive stars.
In Fig. 3b, we subdivide the sample into sub-samples with different C. It is clear that
the the concentration value does not contribute to the dispersion.

%\begin{figure*}
%   \centering
%   \includegraphics[scale=0.8,angle=-90]{fig4.ps}
   %%%\includegraphics{empty.eps}
   %%%\includegraphics{empty.eps}
%   \caption{The relation between SFR and the luminosity combined by 12 \& 22 $\mu$m for our data sample.
%      The SFR is in unit of M$_{\sun}$/yr. The luminosity is in unit of L$_{\sun}$.}
%              \label{Fig4}%
%    \end{figure*}

\subsection{4000$\AA$ break}

To show the contribution of the star formation history to the dispersion
in the relation between SFR and the luminosity at 3.4, 4.6, 12 \& 22 $\mu$m,
 we divided the data sample  into  sub-samples with different 4000$\AA$ breaks  (Fig. 3c).

  We caution that because these 4000$\AA$ break quantities are derived within the fiber aperture, they may not
be representative of the galaxy as a whole, whereas the SFR estimates are for the entire galaxy.
 {\bf If we make the same plot using SFR  within the fiber aperture, we still recover a similar behavior as Fig. 3c which
use the SFR of the entire galaxy. From that, 
we can conclude that the influence of the fiber aperture effect on this relation is negligible. }

It shows that the luminosity-SFR relation at a low 4000$\AA$ break is displaced to higher SFR and lower luminosity.
Brinchmann et al.(2004) show that most of the star formation takes place in galaxies with a low 4000$\AA$ break.
It can explain  the behavior of the low 4000$\AA$ break displaced to higher SFR,
where galaxies with a low 4000$\AA$ break have a younger stellar population, and therefore higher SFR values
than high 4000$\AA$ break galaxies of the same SFR.
The selection effects also contribute to the effect that high 4000$\AA$ break galaxies
tend to be excluded from our data sample because these galaxies usually have old stellar
populations and seldom show evident emission lines. If one high 4000$\AA$ break galaxy indeed shows and
evident emission line, it will tend to have less starbursts, hence the lower SFR.

\subsection{g-z color}

It is interesting to study whether  color relates to  SFR or not. To show this view clearly,
 we divided the data sample  into  sub-samples with different g-z colors (Fig. 3d).

It shows that the luminosity-SFR relation at the bluer end is displaced to higher SFR and lower luminosity.
We can explain this trend like the 4000$\AA$ break where the galaxies with bluer colors
are younger, and therefore have a lower 4000$\AA$ break.  The observed distribution of  the points
in each g-z color bin is linear least-square-fitted by the  following expression,
given as the dashed line in Fig. 3d.

\begin{eqnarray}
\lg(SFR) = -5.70  + 0.65  \lg(L_{\rm W4}) \hspace{0.5cm}  (g-z\leq0.95)  \
\end{eqnarray}

\begin{eqnarray}
\lg(SFR) = -6.10 + 0.65  \lg(L_{\rm W4})  \hspace{0.5cm}(0.95 < g-z\leq1.35)  \
\end{eqnarray}

\begin{eqnarray}
\lg(SFR) = -6.50 + 0.65  \lg(L_{\rm W4})  \hspace{0.5cm}(1.35 < g-z\leq1.95)  \
\end{eqnarray}

\begin{eqnarray}
\lg(SFR) = -8.00  + 0.72   \lg(L_{\rm W4}) \hspace{0.5cm} (g-z  > 1.95)   \
\end{eqnarray}

\section{Conclusions}

We have collected from WIDE-FIELD INFRARED SURVEY (WISE), a large
sample of star-forming galaxies covering a wide range in metallicity
$7.66<\zoh<9.46$, in redshift less than 0.4, and matched the data with
a sample of SDSS DR8 data.
We have found the existence of the correlation between 3.4, 4.6, 12 and
 22 $\mu$m luminosity and SFR and have obtained a calibration that can
be used as a method for determining SFR for star-forming galaxies.
The calibrations hold for galaxies with high quality spectral
observations. The dispersion and non-linearity between 3.4, 4.6, 12 and
 22 $\mu$m luminosity and SFR is related to the galaxy's properties,
such as the 4000$\AA$ break and galaxy color.

%__________________________________________________________________

\section{acknowledgements}

 This work was funded by the National Natural Science Foundation of
China (NSFC) (Grant No.~10873012), the National Basic Research Program
 of China (973 Program) (Grant No.~2007CB815404),  the Chinese
Universities Scientific Fund (CUSF), and the Special fund for Ph.D of
North China Institute of Aerospace Engineering(KY-2010-02-B).

This publication makes use of data products from the Wide-field Infrared
Survey Explorer, which is a joint project of the University of California,
Los Angeles, and the Jet Propulsion Laboratory/California Institute of
Technology, funded by the National Aeronautics and Space Administration.
Funding for the Sloan Digital Sky Survey (SDSS) has been provided
by the Alfred P. Sloan Foundation, the Participating Institutions, the
National Aeronautics and Space Administration, the National Science
Foundation, the U.S. Department of Energy, the Japanese Monbukagakusho,
and the Max Planck Society.

%\end{acknowledgements}

\end{document}